\documentclass[aps,prl,twocolumn,superscriptaddress]{revtex4}

\usepackage{graphicx,amsmath,verbatim,amsfonts} 
\usepackage{textcomp,bbm}

\usepackage{color} \definecolor{blue}{rgb}{0,0,1}

\newcommand{\deff}{d_\mathrm{eff}} 
 \newcommand{\F}{\mathcal{F}}
 \newcommand{\be}{\begin{equation}}
\newcommand{\ee}{\end{equation}}

\begin{document}
	
	\title{Polariton hyperspectral imaging of two-dimensional semiconductor
		crystals}
	
	\author{Christian Gebhardt} \affiliation{Fakult{\"a}t f{\"u}r Physik,
		Ludwig-Maximilians-Universit{\"a}t, Schellingstra{\ss}e~4, 80799~M{\"u}nchen,
		Germany} \affiliation{Max-Planck-Institut f{\"u}r Quantenoptik, 
		Hans-Kopfermann-Str.~1, 85748~Garching, Germany}
	
	\author{Michael F{\"o}rg} \address{Fakult\"at f\"ur Physik, Munich Quantum
		Center, and Center for NanoScience (CeNS), Ludwig-Maximilians-Universit\"at,
		Geschwister-Scholl-Platz 1, 80539 M\"unchen, Germany}
	
	
	\author{Hisato Yamaguchi} \author{Ismail Bilgin} \author{Aditya D. Mohite}
	\affiliation{3MPA-11 Materials Synthesis and Integrated Devices, Materials
		Physics and Applications Division, Los Alamos National Laboratory (LANL), Los
		Alamos, New Mexico 87545, USA}
	
	\author{Christopher Gies} \author{Malte Hartmann} \author{Matthias Florian}
	\affiliation{Institut f{\"u}r Theoretische Physik, Universit{\"a}t Bremen, 28334
		Bremen, Germany} 
	\author{Theodor W. H{\"a}nsch} \affiliation{Fakult{\"a}t
		f{\"u}r Physik, Ludwig-Maximilians-Universit{\"a}t, Schellingstra{\ss}e~4,
		80799~M{\"u}nchen, Germany} \affiliation{Max-Planck-Institut f{\"u}r
		Quantenoptik,  Hans-Kopfermann-Str.~1, 85748~Garching, Germany}
	
	\author{Alexander H{\"o}gele} \address{Fakult\"at f\"ur Physik, Munich Quantum
		Center, and Center for NanoScience (CeNS), Ludwig-Maximilians-Universit\"at,
		Geschwister-Scholl-Platz 1, 80539 M\"unchen, Germany}
	
	\author{David Hunger} \email[To whom correspondence should be addressed. E-mail:
	]{david.hunger@kit.edu} \affiliation{Physikalisches Institut, Karlsruher
		Institut f{\"u}r Technologie, Wolfgang-Gaede-Str.1, 76131 Karlsruhe, Germany}
	
	\date{\today}
	
	\begin{abstract}
		Atomically thin crystals of transition metal dichalcogenides (TMDs) host
		excitons with strong binding energies and sizable light-matter interactions. Coupled to optical cavities, monolayer TMDs routinely reach the regime of strong
		light-matter coupling, where excitons and photons admix coherently to form quasiparticles known as polaritons up to room temperature. Here, we explore the two-dimensional nature of TMD polaritons with cavity-assisted hyperspectral imaging. Using extended WS$_2$ monolayers, we establish the regime of strong coupling with a scanning microcavity to map out polariton properties and correlate their
		spatial features with intrinsic and extrinsic effects. We find a high level of homogeneity, and show that polariton splitting variations are correlated with intrinsic exciton properties such as oscillator strength and linewidth. Moreover, we observe a deviation from thermal equilibrium in the resonant polariton population, which we ascribe to non-perturbative polariton-phonon coupling. Our measurements reveal a promisingly consistent polariton landscape, and highlight the importance of phonons for future polaritonic devices.
	\end{abstract}
	
	

	\maketitle
	
	\textbf{}

	
	Exciton polaritons can enable novel photonic elements such as ultra-low
	threshold lasers \cite{Fraser16,Ye15}, Bose-Einstein condensates \cite{Jiang14},
	or quantum nonlinear optical elements \cite{Verger06}. Atomically thin crystals
	of TMDs offer a particularly promising platform to
	study and harness exciton polaritons due to a strong exciton binding energy
	\cite{Ramasubramaniam12,Chernikov14}  and a large oscillator strength
	\cite{Mak10,Splendiani10,Li14}. Both properties in concert have enabled the
	demonstration of exciton polaritons at room temperature
	\cite{Liu15,Flatten16b,Lundt16} and under cryogenic conditions
	\cite{Dufferwiel15,Wang16,Sidler17}. Owing
	to their unique spin-valley degrees of freedom inherited from the
	non-centrosymmetric host crystal with strong spin-orbit effects
	\cite{Xiao12}, TMD polaritons could enable novel
	photonic devices with topological properties in accordingly
	structured two-dimensional photonic and excitonic landscapes
	\cite{Karzig15}.
	
	It remains an important task to understand the conditions that govern polariton
	properties, and to realize large-scale systems that are useful also for such advanced
	devices. Elevated temperature and the crystals' two-dimensional geometry constitute an environment that strongly influences polariton properties. Variation of both intrinsic defect concentrations and of the dielectric surrounding can lead to a significant spatial variation of the materials optical properties, and phonons couple strongly to the optical transition.
	In this work we study such spatial variations and environmental influences by scanning cavity microscopy in the strong coupling regime. We reveal correlations with intrinsic excitonic properties and observe a polariton population with significant deviations from thermal equilibrium, indicative of a non-Markovian polariton-phonon coupling.
	
\begin{figure}[htb!] \centering
		\includegraphics[width=\columnwidth]{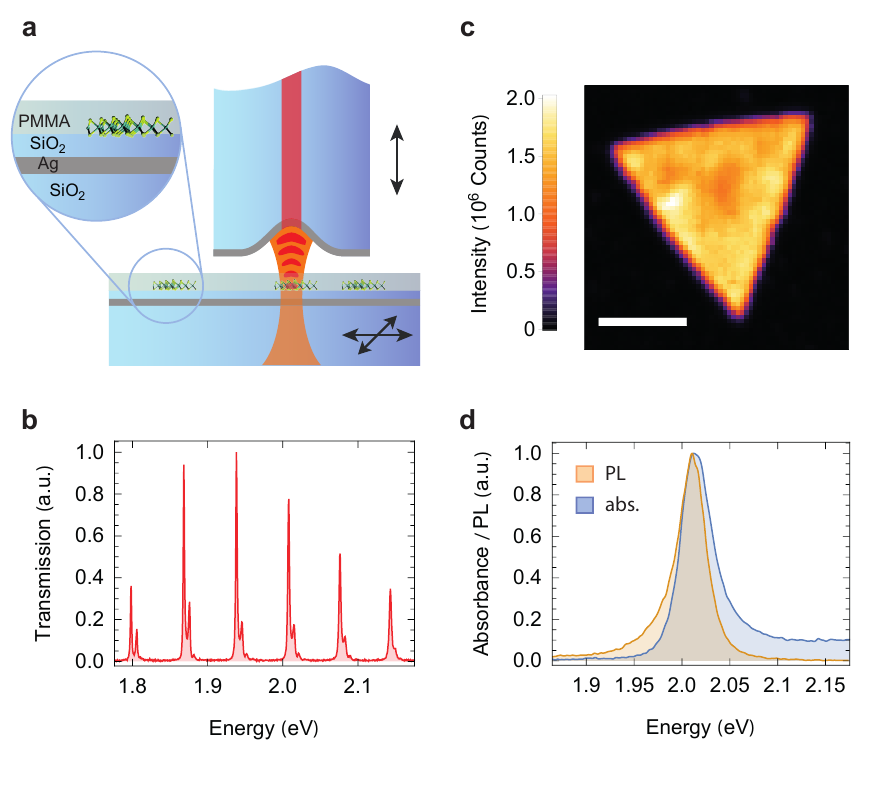}
		\caption{\label{fig:setup} \textbf{Experimental setup and characterisation.} a) A fiber-based microcavity couples to monolayer WS$_2$ covered with a thin film of PMMA on top of a silver (Ag) mirror capped with a dielectric spacer (SiO$_2$). The planar mirror is mounted on a 3D
			nanopositioning stage to enable raster-scanning of the sample through the cavity
			mode and coarse tuning of the mirror separation. The cavity length and thus its
			resonance frequency can be fine-tuned by a piezo actuator moving the fiber. b)
			White light transmission spectrum of the empty cavity for a large mirror
			separation $d = 10.5~\mu$m. c) Confocal photoluminescence map of a WS$_2$
				flake on the mirror (scale bar 10~$\mu$m). d) Confocal photoluminescence (orange) and
				cavity-assisted absorption (blue) spectra recorded on a typical position of a
				WS$_2$ flake on the mirror.} \end{figure}

	Our experimental platform is a fiber-based Fabry-Perot microcavity
	\cite{Hunger10b} consisting of a laser-machined optical fiber serving as a
	micromirror and a planar mirror with monolayer flakes of WS$_2$ synthesized
		by chemical vapor deposition (CVD) and covered with a thin film of PMMA (see
		Fig.~\ref{fig:setup}a and Methods). Away from the flakes, the transmission of
		the bare cavity shown in Fig.~\ref{fig:setup}b features the characteristics of a
		stable Fabry-Perot resonator with Hermite-Gaussian eigenmodes that exhibits a
		strong main resonance and a blue-detuned weak resonance stemming from higher
		transverse modes. The empty cavity has a finesse of 40 at the
	exciton energy of 2.01~eV, leading to a cavity-length dependent linewidth
	$\kappa = 51/q~$meV, where $q$ is the longitudinal mode order.
	
	In the first experimental step, we characterized the CVD-grown WS$_2$ monolayer
	flakes on the macroscopic mirror with confocal microscopy and spectroscopy and
	found extended triangular flakes with bright PL as in Fig.~\ref{fig:setup}c. We observe
	PL spectra as shown in Fig.~\ref{fig:setup}d with a full-width-half-maximum (FWHM) linewidth $\gamma= 33$~meV, typical for monolayer WS$_2$ \cite{Molas17}. For the
	same position, we also measured the absorption spectrum of Fig.~\ref{fig:setup}c
	by cavity-enhanced spectroscopy at large mirror separation where the
	exciton-photon coupling is weak (see methods). The absorption spectrum matches
	the center energy and the linewidth of the emission, but shows additional finite
	background absorption at the higher energy side of the resonance.

	\begin{figure*} \centering \includegraphics{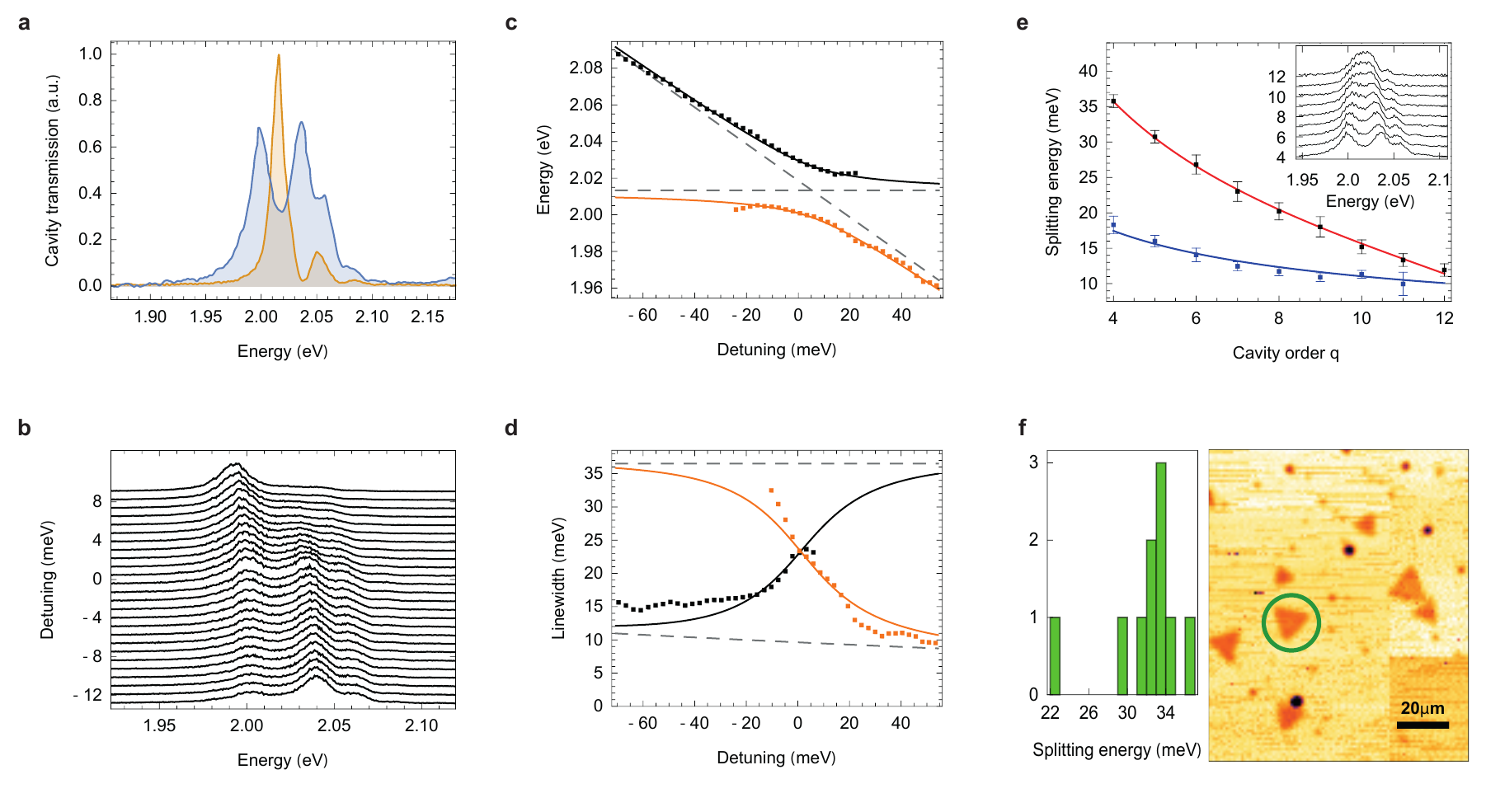}
		\caption{\label{fig:splitting} \textbf{Polariton spectroscopy.} (a) Bare cavity
			transmission spectrum (orange) and transmission spectrum under normal mode
			splitting at mode order $q=4$ (blue). b) Cavity transmission spectra of the
			$q=4$ mode when tuned stepwise across the exciton resonance by varying the mirror separation.
			c) Resonance energies of the upper (black) and lower (orange) polariton
				branches (data) derived from from the spectra in b) and fits to the coupled
				oscillator model (solid lines; dashed lines show the empty cavity and bare
				exciton resonances). d) Extracted linewidths of the upper (black) and lower
				(orange) polariton branches (data) and model fits (solid lines; the upper and
				lower dashed lines show the linewdiths of the empty cavity and the bare exciton
				resonances, respectively). e) Polariton splitting $\hbar\Omega$ (upper data,
			fitted red line, see text) and coupling strength $g$ (lower data and blue line) as a function of cavity mode order $q$ evaluated from the spectra shown in the inset. f)
			Statistics of polariton splitting observed for 10 different flakes (left) and a
			map of cavity-enhanced sample extinction (right).} \end{figure*}

After these initial studies of confocal PL and absorption of
monolayer WS$_2$ flakes at large cavity-mirror separation, we
reduced the intermirror distance to increase the cavity-flake
coupling. For a mirror separation corresponding to the
cavity mode order $q=4$, Fig.~\ref{fig:splitting}a shows the cavity transmission
spectra on (blue) and off (orange) the flake under white light illumination. The
WS$_2$ transmission spectrum exhibits a pronounced normal mode splitting with
two well-resolved polariton resonances in response to the fundamental cavity
mode, and a weak higher-energy resonances due to transverse
modes of the cavity. Tuning the cavity resonance across the emission spectrum
results in avoided level crossing with spectra in Fig.~\ref{fig:splitting}b,
which we compare to a coupled oscillator model in Fig.~\ref{fig:splitting}c and d.
	
	The model accounts for the coupling with strength $g$ between an optical cavity
	with the fundamental mode decay rate $\kappa_0$ and an exciton transition with
	homogeneous linewidth $\gamma$ to obtain the normal-mode or Rabi splitting
	$\Omega$ of the coupled oscillator system \cite{Savona95}: 
	\begin{equation}
	\Omega(q)=2\sqrt{\sqrt{g^4+2g^2\gamma(\gamma+\kappa_0/q)}-\gamma^2}.
	\end{equation} 
	To model the detuning-dependent dispersion of the upper and lower
	polariton modes shown in Fig.~\ref{fig:splitting}c we take $\kappa_0=51~$meV and
	$\gamma=33~$meV as determined from the initial characterization described above
	and use $g$ as the only free fit parameter. Figure~\ref{fig:splitting}e shows
	the observed splitting and the deduced light-matter coupling strength $g$ as a
	function of the longitudinal mode order $q$.
	
	The splitting is smaller than reported in a similar experiment \cite{Flatten16b} (with
	$\hbar\Omega\approx 50\,$meV at $q=4$), which can be traced back to a non-ideal
	PMMA layer thickness that reduces the local field at the sample in our
	experimental configuration. We note, however, that we use a stable microcavity
	where excitons couple to a single, spectrally isolated cavity mode, while
	in \cite{Flatten16b} a planar Fabry-Perot displaying a mode continuum was used.
	
	The model also accounts for the detuning dependence of the
	upper and lower polariton branch linewidths with $\Gamma = (\kappa + \gamma)/2$ on
	resonance (Fig.~\ref{fig:splitting}d). Since the cavity linewidth $\kappa$ is
	significantly narrower than the exciton linewidth $\gamma$, coupling leads to a
	reduced polariton linewidth and thus to an increase of the polariton
	lifetime compared to the exciton lifetime by a factor of 1.7 on resonance, and
	up to a factor 4 at large detuning.
	
	The observations of normal mode splitting and line narrowing as hallmarks of the strong-coupling regime are robust
	characteristics of the light-matter coupling for our sample. The
	right panel of Fig.~\ref{fig:splitting}f shows cavity-enhanced absorption
	measurements of other individual triangular WS$_2$ monolayers. The left panel of
	Fig.~\ref{fig:splitting}f shows the statistics of resonant coupling experiments
	at $q=4$ carried out on 14 of such flakes: while 10 flakes exhibited splitting
	energies in the range of $\hbar \Omega = 22 - 36~$meV, four flakes did not
	exhibit signatures of strong coupling at all. Remarkably, most WS$_2$
	flakes in strong coupling featured comparable coupling strengths with a
	variation in $g$ of only $\pm 3~$meV.

	\begin{figure*} \centering \includegraphics{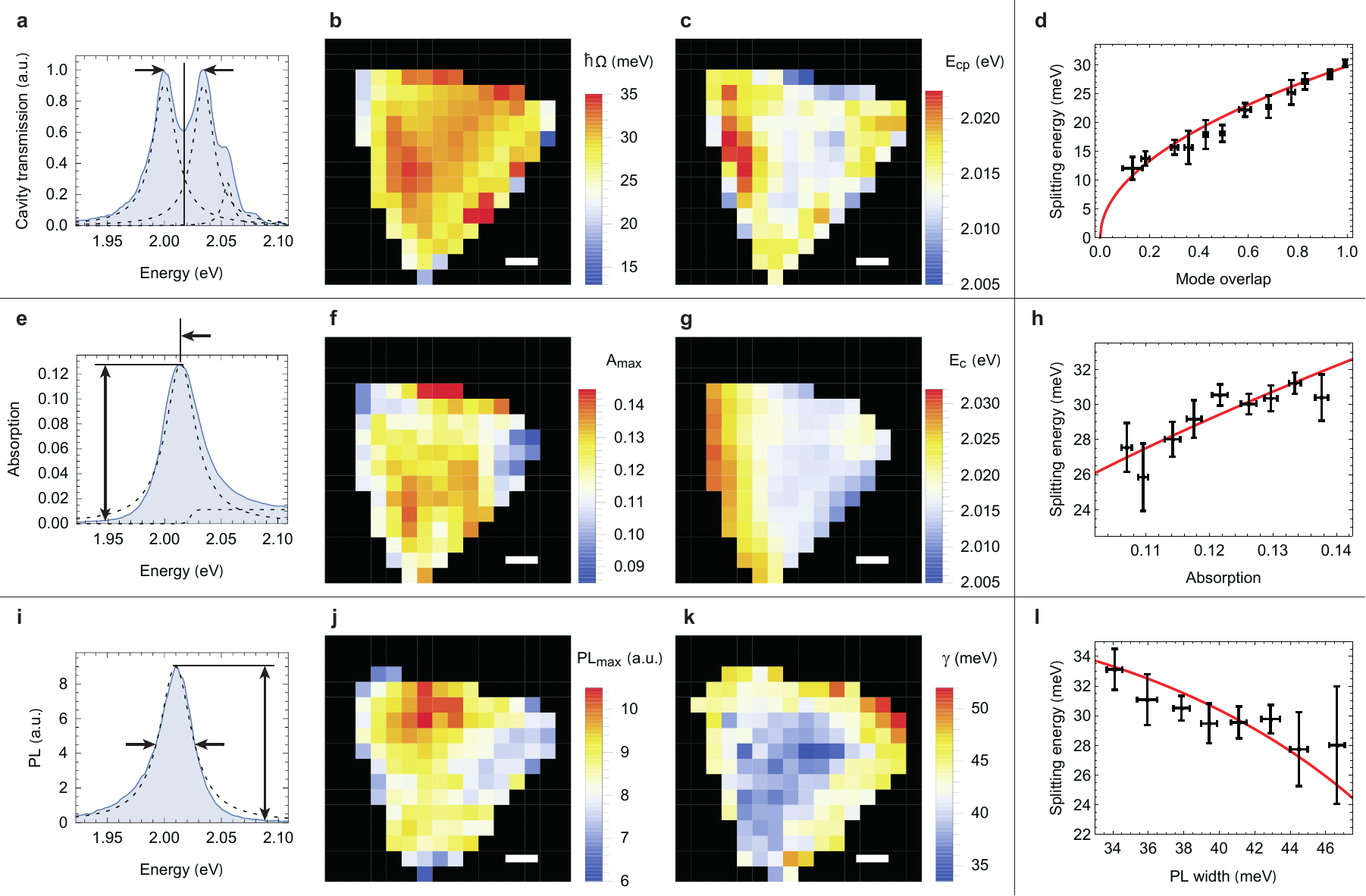}
		\caption{\label{fig:imaging} \textbf{Polariton hyperspectral imaging.} a)
			Resonant cavity transmission showing a normal mode spectrum at $q=4$. Spatial
			maps of  b) the Rabi splitting energy $\hbar\Omega$, and c) the center energy of
			the normal mode spectrum $E_{cp}$. d) Dependence of $\hbar \Omega$ on the
			spatial mode overlap with the flake. e) Absorption spectrum at large $q$ as
			inferred from white light transmission spectroscopy. Maps of f) the maximal
			resonant absorption $A_{max}$, and g) of the absorption center energy $E_c$. h)
			Correlation between $\hbar\Omega$ and the maximum resonant absorption, together
			with a fit (solid line, see text). i)
			Photoluminescence spectrum, j) map of the peak photoluminescence intensity, and
			k) photoluminescence linewdith. l) Correlation between $\hbar\Omega$ and the
			photoluminescence linewidth together with a fit to Eq.~1 (solid line). The scale bar shows 2~$\mu$m in all maps.}
	\end{figure*}
	
	Having established the signatures of strong coupling at individual spatial positions, we utilized the scanning capabilities of our cavity \cite{Mader15,Huemmer16} to study the
	two-dimensional nature of polariton landscapes in extended WS$_2$ flakes. To
	this end, we stabilized the cavity length at $q=4$ and ensure symmetric polariton populations as monitored in	transmission spectroscopy. Under such conditions, the sample was displaced with
	respect to the cavity mode in $1~\mu$m steps corresponding to the spatial
	resolution determined by the cavity mode waist. At each raster-scan pixel we
	recorded white-light transmission spectra of the cavity as in
	Fig.~\ref{fig:imaging}a to infer the polariton splitting, shown in
	Fig.~\ref{fig:imaging}b, and the center energy of the polariton doublet, shown
	in Fig.~\ref{fig:imaging}c.
	
	First, we accounted for the spatial overlap between
	the cavity mode and the flake. Figure~\ref{fig:imaging}d shows how the observed
	Rabi splitting scales with the area overlap $\eta$. The dependence agrees very
	well with the expectation for collective coupling, where $g \sim \sqrt{\eta}$. To eliminate this overlap effect from further analysis, the map of
	Fig.~\ref{fig:imaging}b was corrected to show actual Rabi splitting values
	renormalized as $\Omega = \Omega_{raw}/\sqrt{\eta}$. Similar overlap corrections were performed for the photoluminescence and absorption map Fig.~\ref{fig:imaging} f, j, according to $~1/\eta$, and all data shown is limited to $\eta>0.35$. The remaining
	deviations at edges are consistent with the notion of edges as line defects with
	light-matter characteristics distinct from the flake interior \cite{Neumann17}.
	
	It is instructive to confront the maps of polariton characteristics in the top
	panel of Fig.~\ref{fig:imaging} with complementary parameters of light-matter
	coupling probed for the same flake with resonant absorption (central panel of
	Fig.~\ref{fig:imaging}) and PL (bottom panel of Fig.~\ref{fig:imaging}). The
	absorption spectrum of Fig.~\ref{fig:imaging}e was recorded for a large mirror
	separation ($\sim 10~\mu$m) and used to evaluate the peak absorption (Fig.~\ref{fig:imaging}f)
	and its center energy (Fig.~\ref{fig:imaging}g). 
	For similar conditions, we also record the PL spectrum (Fig.~\ref{fig:imaging}i) and extracted the peak PL
	(Fig.~\ref{fig:imaging}j) and linewidth (Fig.~\ref{fig:imaging}k).
	
	This comprehensive set of spatially-resolved measurements allows us to correlate
	external and internal properties of light-matter interaction in
	monolayer WS$_2$. The polariton splitting in Fig.~\ref{fig:imaging}b shows a
	high degree of homogeneity with moderate variation across the flake, and a
	localized maximum along a line parallel to the left edge. This region is also
	apparent in a blue-shift of the polariton center energy $E_{cp}$
	(Fig.~\ref{fig:imaging}c) and also in the absorption center energy $E_{c}$
	(Fig.~\ref{fig:imaging}g). The transition energy is sensitive to various
	internal and external parameters including strain
	\cite{Conley13,Steinhoff15,He16}, doping \cite{Chernikov15}, or screening due to
	the dielectric environment \cite{Ugeda14,Roesner16,Raja17,Florian17}.
	
	Based on our data, strain is an unlikely explanation since
	it would correlate with the PL yield. However, we do not observe significant
	correlations between PL and $E_c$. We also find no indication of a
	doping-induced energy shift, since no trion contribution was observed in the
	spectra \cite{Plechinger15}. The most probable scenario is that variations in
	the local dielectric environment are responsible for the inhomogeneities in
	light-matter coupling across the flake. The process used to transfer the flakes
	onto the mirror can lead to adsorbates such as water or KOH molecules located
	between the WS$_2$ flake and the SiO$_2$ surface, thereby introducing variations
	in the distance between the WS$_2$ monolayer and the substrate. It is well known
	that the dielectric environment has an extraordinarily large impact on the
	excitonic properties of semiconductor monolayers, causing line shifts of up to
	hundreds of meV \cite{Ugeda14,Roesner16,Raja17,Florian17}. Our calculations show
	that variations of about 1~nm in the interlayer distance result in an energy
	shift of $\sim 10$~meV, consistent with our experimental observations (see
	Supplementary Information).
	
	Comparing the splitting $\hbar\Omega$ with the resonant absorption map $A_{max}$
	shows a clear linear correlation, see Fig.~\ref{fig:imaging}h.  Indeed this follows the expectation for an exciton oscillator strength that is directly proportional to the absorption, evidencing that the observed variations in $A_{max}$ are dominated by radiative excitonic transitions rather than quenching induced by defects. 
	Notably, there is no apparent correlation between the integrated PL and the
	splitting. However, a clear correlation can be observed for the PL linewidth
	(and similarly for the absorption linewidth, not shown). As expected from Eq.~1,
	an anti-correlation between $\Omega$ and $\gamma$ is expected. Originating
	either in excessive pure dephasing or inhomogeneous broadening, the increased
	linewidth towards the edges of the flake indicate an increased inhomogeneity and
	dissipation at these locations. Figure~\ref{fig:imaging}l shows good agreement
	between the observed linewidth dependence of the Rabi splitting and Eq.~1. 
	Since the PL linewidth and the absorption center
	energy remain rather constant in the area where the absorption displays its
	largest variation, the different dependencies can be disentangled to a high
	degree, and the correlations shown in Fig.~\ref{fig:imaging}h,l) are dominated
	by a single quantity.
	Overall, we find that the variation in polariton splitting is governed by
	spatial variation of resonant absorption, inhomogeneous broadening and excessive
	pure dephasing or dissipation. With this we can link the polariton splitting to intrinsic exciton properties in a quantitative way.
	
\begin{figure} \centering \includegraphics{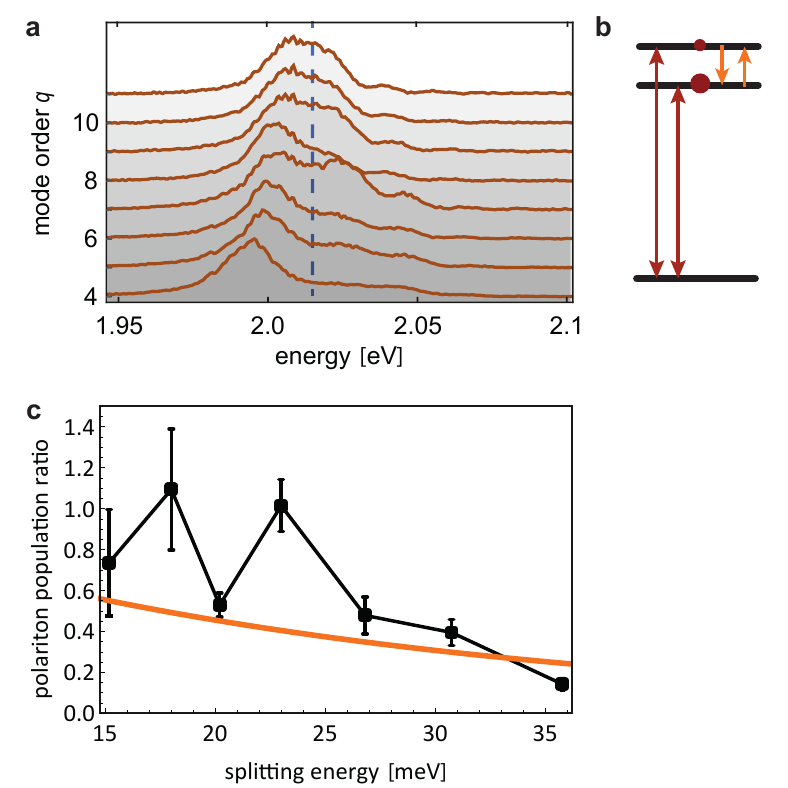}
		\caption{\label{fig:phonons} \textbf{Polariton
				asymmetry due to phonon transitions.} a) Normal mode spectra for different mode
			orders $q=4 - 11$ with the cavity tuned to exact resonance with the exciton. b)
			Schematic level scheme of the ground state and two polariton states with optical
			(red) and phonon (orange) transitions. c) Polariton population ratio as a
			function of polariton splitting energy. The solid line shows the Boltzmann
			distribution.} \end{figure}

	Finally, we turn to the influence of phonons on the polariton spectrum. It was
	suggested that phonons can lead to a surprising departure from a semi-classical
	description of the polaritons \cite{IlesSmith16,Chovan08}. The directly
	observable effect is that the phonon bath mediates direct transitions between
	the dressed states and thereby affects the polariton population. In the
	experiment, we inspect the polariton population for exact cavity resonance
	conditions, i.e. where the polariton splitting is minimal for a given mode order
	$q$. Figure~\ref{fig:phonons}a shows the spectra for different $q$. A large
	asymmetry is apparent, which increases non-monotonically for decreasing cavity
	mode order. We do not expect other processes such as exciton-electron scattering
	or exciton-exciton scattering to be relevant for our experimental conditions
	since we observe no signs of trions and excite with sufficiently low powers such
	that the exciton population remains small. We evaluate the population ratio $p_u/p_l$ of
	the upper and lower polariton resonance from fits to the spectra and display the
	result as a function of the polariton splitting energy, see
	Fig.~\ref{fig:phonons}c. In this way, we can perform spectroscopy of the
	inter-polariton scattering strength within a single device at a fixed
	temperature. This is advantageous since the sample properties and the phonon
	bath remain fixed in this way. We find that the overall behavior is consistent
	with a thermal occupation following a Boltzmann distribution \cite{Michetti09}
	$p_u/p_l = \exp{-\hbar\Omega/k_B T}$ (solid line in Fig.~\ref{fig:phonons}b, no
	free parameters), but with a significant deviation in particular for mode order
	$q=7$ ($\hbar\Omega=23$meV) and $q=9$ ($\hbar\Omega=18$meV), where the upper
	polariton remains significantly more populated than for a thermalized
	distribution.
	
	Such a deviation could originate from a phonon bottleneck, resulting from a low
	phonon density of states at the relevant energy, such that thermalization is
	suppressed. However, the splitting energy of mode $q=7$ ($q=9$) coincides with
	the energy of longitudinal acoustic (transverse acoustic) phonons at the K point
	with a significant density of states \cite{MolinaSanchez11}. Indeed it has been
	shown that a large exciton-phonon coupling strength is present under resonant
	conditions by exciton luminescence up-conversion \cite{Jones15}. The apparent
	impact of K-point phonons here suggests that transitions between polaritons
	states are accompanied by inter-valley scattering. The resonant polariton-phonon
	coupling establishes a non-thermal distribution indicative of non-Markovian
	dynamics \cite{Chovan08,deLiberato09,Michetti09}. Such a scenario is expected to
	lead to superposition states involving polariton, valley, and phonon degrees of
	freedom \cite{Chovan08} with emergent quantum correlations \cite{IlesSmith16}.
	
	Our work shows that spatial variations in the optical properties of atomically
	thin semiconductors and their local environment influence the polariton states
	of a coupled monolayer-microcavity system and lead to variations in the
	polariton splitting on a few-micron scale within a flake, as well as for
	different flakes. Clear correlations were established interconnecting local
	variations in the oscillator strength and the exciton linewidth with the Rabi
	splitting. While the observed variations are already encouragingly small,
	further improvement in homogeneity is expected for flakes embedded in van der
	Waals heterostructures e.g. with hexagonal boron nitride. This could open the
	way for advanced polaritonic devices such as tunable polariton lasers and
	polariton Bose-Einstein condensation, and enable the realization of topological
	polaritons \cite{Karzig15}. Furthermore, the resonant interaction of polaritons with
	phonons leads to a significant alteration of the polariton population, with
	marked deviations from a thermal distribution. While further studies are
	required to elucidate this behavior, the observed strong polariton-phonon
	coupling could provide a novel resource for polaritonic devices.

	\subsection{Methods} \subsubsection{Cavity characterization} The mirrors were
	prepared by evaporating 50~nm silver as reflective layer and a 20~nm (100~nm)
		SiO$_2$ capping layer on top which protects silver from oxidation and serves as
		a spacer layer to place the sample at a field antinode. The fiber, shaped by
	CO$_2$ laser machining \cite{Hunger12}, has a conical tip to achieve smallest
	mirror separations \cite{Kaupp16}. In its center, we fabricate a concave profile
	with a radius of curvature of $75~\mu$m and a depth of $200~$nm. We measure the
	cavity finesse $\F$ with a laser at 532~nm and obtain a value $\F=30$, and
	$\F=40$ from white light transmission spectra at the exciton emission energy
	$E_0=2.01$~eV, in good agreement with a simulation. To calibrate the optical
	cavity length, we record broad-band cavity transmission spectra with a
	supercontinuum laser and evaluate the separation of subsequent cavity
	resonances, see Fig.~\ref{fig:setup}b. We find that the smallest accessible
	effective cavity length is $\deff=4\lambda/2$, corresponding to the longitudinal
	mode order $q = 4$, limited by the profile depth (200~nm) and the presence of
	the PMMA layer which covers the sample. At this separation, we obtain a cavity
	quality factor of $ Q_c = q \F = 200$. From scanning-cavity microscopy
	measurements and calculations, we infer the mode waist to be $ w_0 = 1.0~\mu$m,
	which approximately defines the spatial resolution of the scanning cavity
	microscope.
	
	\subsubsection{Sample preparation} WS$_2$ monolayer crystals were grown by
	sulfurization of tungsten dioxide (WO$_2$) powder. A SiO$_2$/Si substrate along
	with a WO$_2$ powder boat were placed at the center of a chemical vapor
	deposition (CVD) furnace. The SiO$_2$/Si substrate was facing down in close
	proximity to the WO$_2$ powder ($99.99 \%$, Sigma Aldrich). The temperature was
	initially ramped up rapidly but slowed down to $3 ^\circ$C /min as it approached
	to $850^\circ$C. Sulfur powder ($99.5\%$, Alfa Aesar) was placed at upstream end
	of the quartz tube in a separate boat near the heating zone to allow
	vaporization ($\sim 110 ^\circ$C) during the growth. The growth temperature was
	maintained for 15 minutes before cooling it down to the room temperature. 200
	SCCM of Argon was used as a carrier gas during the entire process. As-grown
	monolayer crystals were studied in spectroscopy or transferred onto a mirror
	using established polymer-supported wet method. To this end polymethyl
	methacrylate (PMMA) was spin-coated on the monolayer flakes and lifted off in 1M
	potassium hydroxide (KOH) in water. The PMMA-supported film with WS$_2$ crystal
	flakes was rinsed in water for three cycles at room temperature to remove
	possible KOH residue and finally transferred onto mirror substrates.
	
	\subsubsection{Photluminescence microscopy and spectroscopy} Excitation was
	performed either with a cw laser at 532nm for PL or with a pulsed supercontinuum
	(Fianium Whitelase 450~SC, 20~MHz, $\sim 50~$ps) filtered to a spectral band of
	580~nm to 650~nm. Confocal measurements were performed in a homebuilt setup
	including a 0.9~NA air objective. Detection was performed either with a Si
	avalanche photodetector or with a grating spectrometer (Princeton Instruments,
	Acton 2500) equipped with a sensitive CCD camera (Andor ikon-M). To observe the
	polariton spectrum we perform broad-band transmission spectroscopy of the
	coupled cavity-emitter system. We reduce the laser power such that during the
	pulse, much less than one photon populates the cavity on average to avoid
	multi-photon processes. The transmitted light was spectrally filtered, fiber
	coupled and recorded with the grating spectrometer.
	
	\subsubsection{Model for environmental exciton energy renormalization} To assess
	the impact of variations in the distance between the optically active WS$_2$
	flake and the SiO$_2$ substrate, we employ a multiscale approach introduced in
	Ref. \cite{Florian17}. In a first step, an electrostatic model is used to
	determine an effective non-local dielectric function to account for screening of
	Coulomb interaction between carriers in the TMD in a vertical heterostructure
	environment. In a second step, band-structure renormalizations and the screened
	Coulomb potential enter calculations of the optical properties to determine the
	spectral positions of the excitonic resonances.
	
	\subsection{Acknowledgments} We thank Thomas H{\"u}mmer, Julia Benedikter,
	Matthias Mader and Hanno Kaupp for support. Fruitful discussions with Jonathan
	Finley and Michael Kaniber are acknowledged. The work has been funded by the DFG
	Cluster of Excellence NIM, the Volkswagen Foundation, and the European Research
	Council (ERC) under the ERC grant agreement no. $336749$. T. W. H\"ansch
	acknowledges funding from the Max-Planck Foundation.
	
	\subsection{Author contributions} D.H. and A.H. conceived the
	project, C.G. and M.F. designed and set up the experiment, H.Y., I.B. and A. D.
	M. prepared the sample, C.G. and M.F. collected the data, C.G., A.H., C. Gies,
	M. Florian, M.H., and D.H. analyzed and modeled the data. C.G., A.H. and D.H. wrote
	the manuscript. All authors contributed to the discussion of the results and the
	manuscript.

	\section{Supplementary Information}

	\section{Setup}
		\begin{figure*}[tb!]
		\centering
		\includegraphics{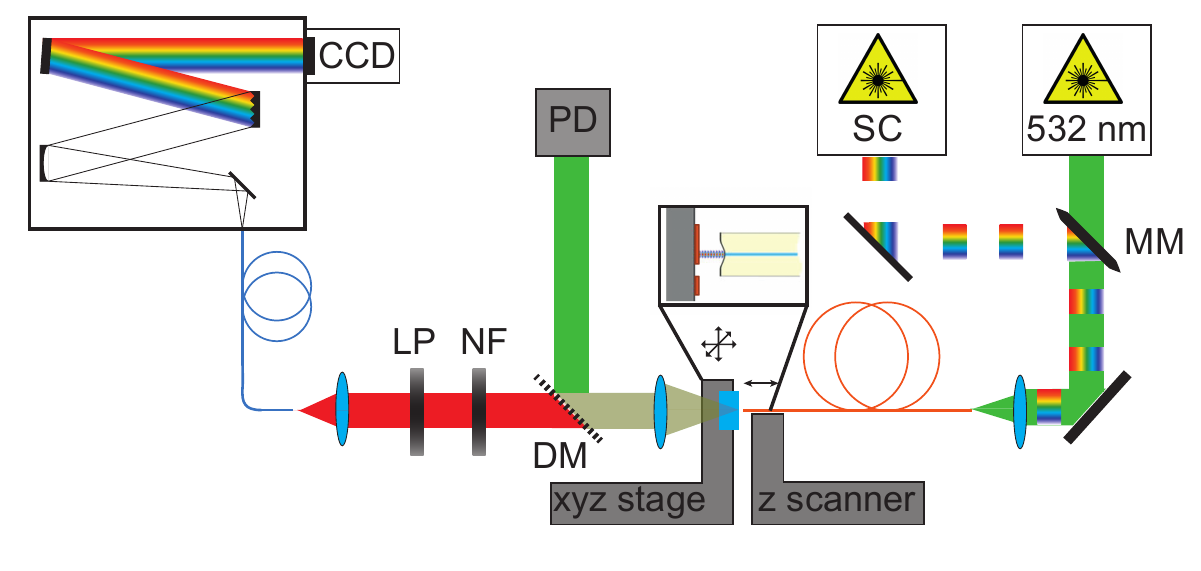}
		\caption{\label{fig:Setup} Schematic drawing of the setup.
		}
	\end{figure*}

	Two laser sources can be coupled into the cavity fiber: a pulsed super-continuum source (SC) and a 532nm cw laser which can be selected via a motorized mirror (MM). The fiber can be moved in z-direction with sub-nm precision via a piezo actuator, while the sample is mounted on an attocube nanopositioner (ECS3030) with three translational degrees of freedom.
	The collected light is split into two paths with a dichroic mirror (DM). The short wavelength part is detected with a photodiode (PD), the long wavelengths pass a notch filter (NF) and a longpass filter (LP) to be detected with a spectrometer consisting of different gratings and a CCD camera.

	\section{Evaluation of cavity transmission spectra}\label{splittingLength}
	
	\subsection{Cavity-enhanced absorption spectroscopy}
	We probe the transmission of broadband light from the SC source through the cavity with the grating spectrometer. We tune a set of cavity resonances stepwise across the spectrum and evaluate the resonances' peak transmission of multiple exposures to obtain a continuous spectrum. We perform measurements with a WS$_2$ flake located in the cavity mode ($I_\mathrm{flk}$) and for an empty cavity as a reference ($I_\mathrm{ref}$). We normalize the transmission spectrum to obtain the cavity-enhanced loss, $B_\mathrm{max}=1-I_\mathrm{flk}/I_\mathrm{ref}$, and calculate the peak absorption $A_\mathrm{max}$ from $B_\mathrm{max}$ and the mirror reflectivity $R$ as obtained from empty cavity measurements by solving the Fabry-Perot transmission function for $A_\mathrm{max}$,
	\begin{eqnarray*}
	A_\mathrm{max}=\frac{2B_\mathrm{max}(R(1-R)+(1-R^2)}{2R^2(1-B_\mathrm{max})} + \\
	\frac{\sqrt{(1-R^2)^2-4B_\mathrm{max}R(1-R)^2}}{2R^2(1-B_\mathrm{max})}.
	\end{eqnarray*}
	This yields calibrated absorption spectra as shown in the manuscript.
	
	\begin{figure}
		\centering
		\includegraphics{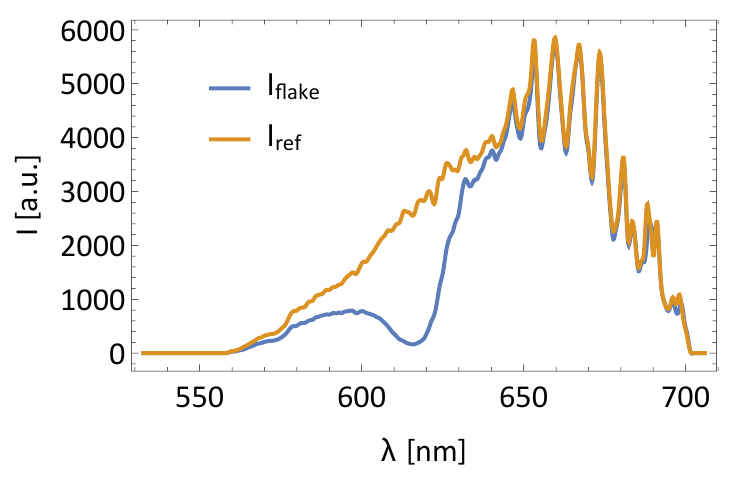}
		\caption{\label{fig:AbsSpec} Broadband cavity-enhanced transmission spectra taken for an empty cavity (orange) and with a WS$_2$ flake (blue).
		}
	\end{figure}

	\subsection{Fitting of transmission spectra}
	We fit all spectra showing normal mode splitting with a triple Lorentzian, accounting for the two polariton resonances and one additional higher order transverse mode of the cavity associated with the upper polariton. The transversal mode of the lower polariton is neglected in the analysis, since its peak is not resolvable in the spectrum, as it is the case for other higher order modes. We expect an analogous anti-crossing behavior of the transversal mode as for the fundamental mode. From this we estimate that the transversal mode of the lower polariton contains less than 1\% of the ground mode polariton intensities. We fit the following expression to the measured spectra:
	\begin{eqnarray*}
	\label{tripleLorentz}
	S(\lambda)& = & a_{up}\frac{\Delta_{up}^2}{(\lambda-\lambda_{up})^2-\Delta_{up}^2} +\\
	 & & a_{lp}\frac{\Delta_{lp}^2}{(\lambda-\lambda_{lp})^2-\Delta_{lp}^2} +\\
	& &  a_{tm}\frac{\Delta_{tm}^2}{(\lambda-\lambda_{tm})^2-\Delta_{tm}^2}
	\end{eqnarray*}
	From the fits we extract the polariton wavelengths \(\lambda_{up},\lambda_{lp}\), the full-widths at half maximum of the polariton branches \(\Delta_{up},\Delta_{lp}\), and the amplitudes \(a_{up},a_{lp}\). Care is taken to obtain optimal fitting, and only reasonable fit values are used for later evaluations.

	\subsection{Polariton dispersion}
	The complex polariton eigenfrequencies are given by the following expression \cite{Savona95}:
	\begin{equation*}
	\omega=\frac{\omega_{ex}+\omega_c}{2} - i\frac{\gamma+\kappa}{4}\pm\sqrt{g^2 + \frac{1}{4}\left(\omega_{ex}-\omega_c-i\frac{\gamma-\kappa}{2}\right)^2}.
	\label{savonaDispersion}
	\end{equation*}
	The dispersion of the polaritons is given by the real part of the spectrum \(Re(\omega)\).
	A simultaneous fit of the two branches to the dispersion data is performed to extract the coupling constant $g$ and other parameters. We perform several fits and fix different parameters ((i) exciton and cavity linewidth, (ii) exciton energy and exciton and cavity linewidth, (iii) energy of zero detuning, exciton and cavity linewidth) in each case to obtain an estimate for the uncertainty of the fit.
	
	\subsection{Polariton width}
	The polariton linewidths are given by the imaginary part of the spectrum \(-Im(\omega)\).
	A simultaneous fit of the linewidths to the two branches is performed to define the exciton linewidth \(\gamma\). As for the dispersion relation, different parameters are fixed ((i) cavity linewidth, (ii) exciton energy and cavity linewidth, (iii) energy of zero detuning and cavity linewidth) to check the robustness of the fits.
	From the models, we obtain an exciton linewidth of \(\gamma\approx 37~\)meV, in good agreement with confocal measurements and from cavity-enhanced fluorescence spectroscopy at large mirror separation. 
	
	\subsection{Polariton intensity}
	We account for the wavelength dependence of the SC source and of the background absorption of the sample when evaluating the polariton intensities. Therefore, the transmission spectra are normalized by the SC spectrum as measured after the cavity without a sample to include wavelength dependent coupling efficiencies (see section above on cavity-enhanced absorption spectroscopy).
	In a second step, the effect of background absorption is corrected by normalizing with the wavelength dependent intracavity loss (not including the resonant absorption from the exciton).
	The normalized polariton intensities are then fitted with

	\begin{eqnarray*}
	p_{u,l} & = & \Theta(\mp\delta)\cos{\left[\frac{1}{2}\arctan{\left(2\frac{g}{\delta}\right)}\right]}^2 +\\  & & \Theta(\pm\delta)\sin{\left[\frac{1}{2}\arctan{\left(2\frac{g}{\delta}\right)}\right]}^2
	\end{eqnarray*}

	for the upper and lower polaritons.
	
	\section{Theoretical model of dielectric screening}
\begin{figure*}
		\centering
		\includegraphics[width=0.8\textwidth]{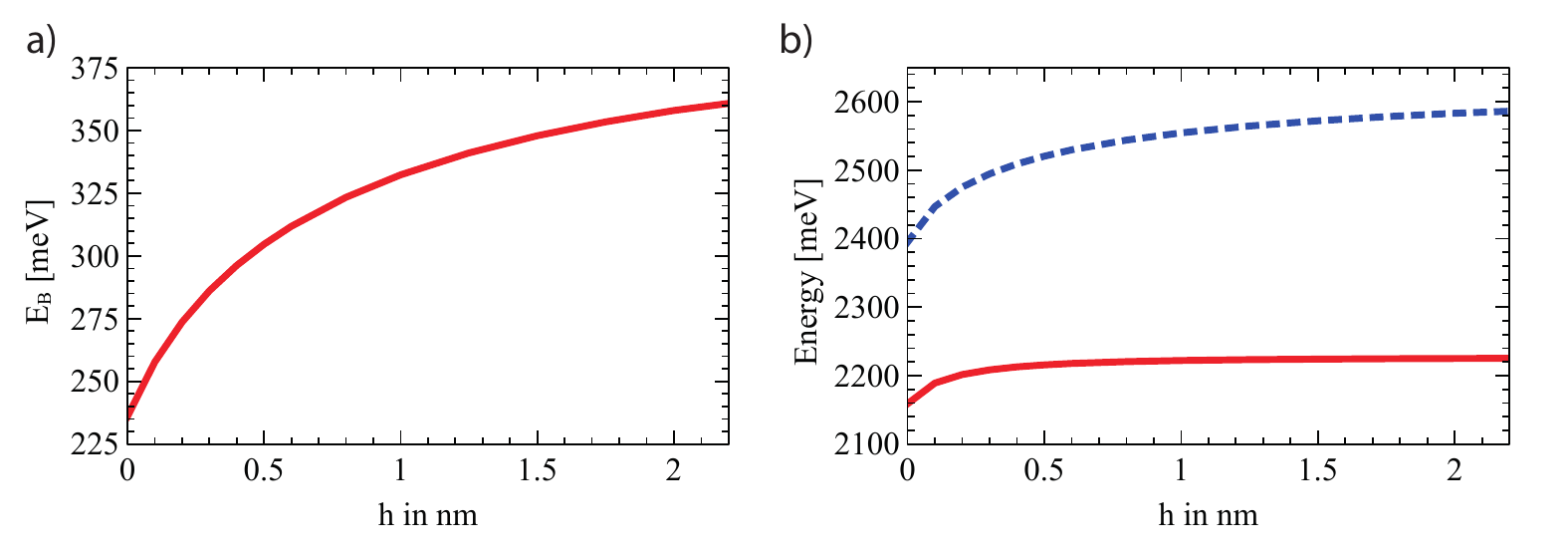}
		\caption{\label{fig:Theory} a) Exciton binding energy as a function of the distance $h$ between the WS$_2$ flake and the SiO$_2$ substrate. b) Resulting gap-dependent center energy of the excitonic absorption peak (red line) and electronic band gap (blue dotted line).
		}
	\end{figure*}

	In Fig.\ref{fig:Theory}, we quantify the effect that local variations in the interlayer distance can have on the optical properties of a WS$_2$ monolayer between a PMMA layer and a silicon oxide substrate. Shown is the dependence on the distance between the TMD and the SiO$_2$ substrate, while the distance between PMMA and TMD is fixed at a realistic value of 5$\AA$ \cite{Rooney17}. The top panel shows the dependence of the band gap at the K point and the exciton binding energy. The resulting impact on the spectral position of the A-exciton transition is shown in the bottom panel. A variation of the gap to silicon oxide between 0.4 nm and 1.4 nm leads to a shift of the excitonic absorption peak of about 11 meV, which agrees with the variation observed in Fig. 3g in the main text.

\begin{thebibliography}{10}
	\expandafter\ifx\csname url\endcsname\relax
	\def\url#1{\texttt{#1}}\fi
	\expandafter\ifx\csname urlprefix\endcsname\relax\def\urlprefix{URL }\fi
	\providecommand{\bibinfo}[2]{#2}
	\providecommand{\eprint}[2][]{\url{#2}}
	
	\bibitem{Fraser16}
	\bibinfo{author}{Fraser, M.~D.}, \bibinfo{author}{H{\"o}fling, S.} \&
	\bibinfo{author}{Yamamoto, Y.}
	\newblock \bibinfo{title}{Physics and applications of exciton-polariton
		lasers}.
	\newblock \emph{\bibinfo{journal}{Nature Materials}}
	\textbf{\bibinfo{volume}{15}}, \bibinfo{pages}{1049} (\bibinfo{year}{2016}).
	
	\bibitem{Ye15}
	\bibinfo{author}{Ye, Y.} \emph{et~al.}
	\newblock \bibinfo{title}{Monolayer excitonic laser}.
	\newblock \emph{\bibinfo{journal}{Nature Photonics}}
	\textbf{\bibinfo{volume}{9}}, \bibinfo{pages}{733} (\bibinfo{year}{2015}).
	
	\bibitem{Jiang14}
	\bibinfo{author}{Jiang, J.-H.} \& \bibinfo{author}{John, S.}
	\newblock \bibinfo{title}{Photonic architectures for equilibrium
		high-temperature bose-einstein condensation in dichalcogenide monolayers}.
	\newblock \emph{\bibinfo{journal}{Scientific Reports}}
	\textbf{\bibinfo{volume}{4}}, \bibinfo{pages}{7432} (\bibinfo{year}{2014}).
	
	\bibitem{Verger06}
	\bibinfo{author}{Verger, A.}, \bibinfo{author}{Ciuti, C.} \&
	\bibinfo{author}{Carusotto, I.}
	\newblock \bibinfo{title}{Polariton quantum blockade in a photonic dot}.
	\newblock \emph{\bibinfo{journal}{Phys. Rev. B}} \textbf{\bibinfo{volume}{73}},
	\bibinfo{pages}{193306} (\bibinfo{year}{2006}).
	
	\bibitem{Ramasubramaniam12}
	\bibinfo{author}{Ramasubramaniam, A.}
	\newblock \bibinfo{title}{Large excitonic effects in monolayers of molybdenum
		and tungsten dichalcogenides}.
	\newblock \emph{\bibinfo{journal}{Phys. Rev. B}} \textbf{\bibinfo{volume}{86}},
	\bibinfo{pages}{115409} (\bibinfo{year}{2012}).
	
	\bibitem{Chernikov14}
	\bibinfo{author}{Chernikov, A.} \emph{et~al.}
	\newblock \bibinfo{title}{Exciton binding energy and nonhydrogenic Rydberg
		series in monolayer ${\mathrm{WS}}_{2}$}.
	\newblock \emph{\bibinfo{journal}{Phys. Rev. Lett.}}
	\textbf{\bibinfo{volume}{113}}, \bibinfo{pages}{076802}
	(\bibinfo{year}{2014}).
	
	\bibitem{Mak10}
	\bibinfo{author}{Mak, K.~F.}, \bibinfo{author}{Lee, C.}, \bibinfo{author}{Hone,
		J.}, \bibinfo{author}{Shan, J.} \& \bibinfo{author}{Heinz, T.~F.}
	\newblock \bibinfo{title}{Atomically thin ${\mathrm{MoS}}_{2}$: A new
		direct-gap semiconductor}.
	\newblock \emph{\bibinfo{journal}{Phys. Rev. Lett.}}
	\textbf{\bibinfo{volume}{105}}, \bibinfo{pages}{136805}
	(\bibinfo{year}{2010}).
	
	\bibitem{Splendiani10}
	\bibinfo{author}{Splendiani, A.} \emph{et~al.}
	\newblock \bibinfo{title}{Emerging photoluminescence in monolayer mos2}.
	\newblock \emph{\bibinfo{journal}{Nano Letters}} \textbf{\bibinfo{volume}{10}},
	\bibinfo{pages}{1271--1275} (\bibinfo{year}{2010}).
	
	\bibitem{Li14}
	\bibinfo{author}{Li, Y.} \emph{et~al.}
	\newblock \bibinfo{title}{Measurement of the optical dielectric function of
		monolayer transition-metal dichalcogenides: ${\mathrm{mos}}_{2}$,
		$\mathrm{Mo}\mathrm{S}{\mathrm{e}}_{2}$, ${\mathrm{ws}}_{2}$, and
		$\mathrm{WS}{\mathrm{e}}_{2}$}.
	\newblock \emph{\bibinfo{journal}{Phys. Rev. B}} \textbf{\bibinfo{volume}{90}},
	\bibinfo{pages}{205422} (\bibinfo{year}{2014}).
	
	\bibitem{Liu15}
	\bibinfo{author}{Liu, X.} \emph{et~al.}
	\newblock \bibinfo{title}{Strong light-matter coupling in two-dimensional
		atomic crystals}.
	\newblock \emph{\bibinfo{journal}{Nature Photonics}}
	\textbf{\bibinfo{volume}{9}}, \bibinfo{pages}{30} (\bibinfo{year}{2015}).
	
	\bibitem{Flatten16b}
	\bibinfo{author}{He, L.~C. F.~Z.} \emph{et~al.}
	\newblock \bibinfo{title}{Room-temperature exciton-polaritons with
		two-dimensional {WS}$_2$}.
	\newblock \emph{\bibinfo{journal}{Scientific Reports}}
	\textbf{\bibinfo{volume}{6}}, \bibinfo{pages}{33134} (\bibinfo{year}{2016}).
	
	\bibitem{Lundt16}
	\bibinfo{author}{Lundt, N.} \emph{et~al.}
	\newblock \bibinfo{title}{Room-temperature tamm-plasmon exciton-polaritons with
		a {WSe}$_2$ monolayer}.
	\newblock \emph{\bibinfo{journal}{Nature Communications}}
	\textbf{\bibinfo{volume}{7}}, \bibinfo{pages}{13328} (\bibinfo{year}{2016}).
	
	\bibitem{Dufferwiel15}
	\bibinfo{author}{Dufferwiel, S.} \emph{et~al.}
	\newblock \bibinfo{title}{Exciton-polaritons in van der Waals heterostructures
		embedded in tunable microcavities}.
	\newblock \emph{\bibinfo{journal}{Nature Communications}}
	\textbf{\bibinfo{volume}{6}}, \bibinfo{pages}{8579} (\bibinfo{year}{2015}).
	
	\bibitem{Wang16}
	\bibinfo{author}{Wang, Q.} \emph{et~al.}
	\newblock \bibinfo{title}{Direct observation of strong light-exciton coupling
		in thin WS$_2$ flakes}.
	\newblock \emph{\bibinfo{journal}{Opt. Express}} \textbf{\bibinfo{volume}{24}},
	\bibinfo{pages}{7151} (\bibinfo{year}{2016}).
	
	\bibitem{Sidler17}
	\bibinfo{author}{Sidler, M.} \emph{et~al.}
	\newblock \bibinfo{title}{Fermi polaron-polaritons in charge-tunable atomically
		thin semiconductors}.
	\newblock \emph{\bibinfo{journal}{Nature Physics}}
	\textbf{\bibinfo{volume}{13}}, \bibinfo{pages}{255} (\bibinfo{year}{2017}).
	
	\bibitem{Xiao12}
	\bibinfo{author}{Xiao, D.}, \bibinfo{author}{Liu, G.-B.},
	\bibinfo{author}{Feng, W.}, \bibinfo{author}{Xu, X.} \& \bibinfo{author}{Yao,
		W.}
	\newblock \bibinfo{title}{Coupled spin and valley physics in monolayers of
		{M}o{S}$_2$ and other group-{IV} dichalcogenides}.
	\newblock \emph{\bibinfo{journal}{Phys. Rev. Lett.}}
	\textbf{\bibinfo{volume}{108}}, \bibinfo{pages}{196802}
	(\bibinfo{year}{2012}).
	
	\bibitem{Karzig15}
	\bibinfo{author}{Karzig, T.}, \bibinfo{author}{Bardyn, C.-E.},
	\bibinfo{author}{Lindner, N.~H.} \& \bibinfo{author}{Refael, G.}
	\newblock \bibinfo{title}{Topologiacl polaritons}.
	\newblock \emph{\bibinfo{journal}{Physical Review X}}
	\textbf{\bibinfo{volume}{5}}, \bibinfo{pages}{031001} (\bibinfo{year}{2015}).
	
	\bibitem{Hunger10b}
	\bibinfo{author}{Hunger, D.} \emph{et~al.}
	\newblock \bibinfo{title}{Fiber Fabry-Perot cavity with high finesse}.
	\newblock \emph{\bibinfo{journal}{New J. Phys.}} \textbf{\bibinfo{volume}{12}},
	\bibinfo{pages}{065038} (\bibinfo{year}{2010}).
	
	\bibitem{Molas17}
	\bibinfo{author}{Molas, M.~R.} \emph{et~al.}
	\newblock \bibinfo{title}{The optical response of monolayer{,} few-layer and
		bulk tungsten disulfide}.
	\newblock \emph{\bibinfo{journal}{Nanoscale}} \textbf{\bibinfo{volume}{9}},
	\bibinfo{pages}{13128--13141} (\bibinfo{year}{2017}).
	
	\bibitem{Savona95}
	\bibinfo{author}{Savona, V.}, \bibinfo{author}{Andreani, L.~C.},
	\bibinfo{author}{Schwendimann, P.} \& \bibinfo{author}{Quattropani, A.}
	\newblock \bibinfo{title}{Quantum well excitons in semiconductor microcavities:
		Unified treatment of weak and strong coupling regimes}.
	\newblock \emph{\bibinfo{journal}{Solid State Communications}}
	\textbf{\bibinfo{volume}{93}}, \bibinfo{pages}{733--739}
	(\bibinfo{year}{1995}).
	
	\bibitem{Mader15}
	\bibinfo{author}{Mader, M.}, \bibinfo{author}{Reichel, J.},
	\bibinfo{author}{H{\"a}nsch, T.~W.} \& \bibinfo{author}{Hunger, D.}
	\newblock \bibinfo{title}{A scanning cavity microscope}.
	\newblock \emph{\bibinfo{journal}{Nature Communications}}
	\textbf{\bibinfo{volume}{6}}, \bibinfo{pages}{7249} (\bibinfo{year}{2015}).
	
	\bibitem{Huemmer16}
	\bibinfo{author}{H{\"u}mmer, T.} \emph{et~al.}
	\newblock \bibinfo{title}{Cavity-enhanced raman microscopy of individual carbon
		nanotubes}.
	\newblock \emph{\bibinfo{journal}{Nature Communications}}
	\textbf{\bibinfo{volume}{7}}, \bibinfo{pages}{12155} (\bibinfo{year}{2016}).
	
	\bibitem{Neumann17}
	\bibinfo{author}{Neumann, A.} \emph{et~al.}
	\newblock \bibinfo{title}{Opto-valleytronic imaging of atomically thin
		semiconductors}.
	\newblock \emph{\bibinfo{journal}{Nature Nanotechnol.}}
	\textbf{\bibinfo{volume}{12}}, \bibinfo{pages}{329--334}
	(\bibinfo{year}{2017}).
	
	\bibitem{Conley13}
	\bibinfo{author}{Conley, H.~J.} \emph{et~al.}
	\newblock \bibinfo{title}{Bandgap engineering of strained monolayer and bilayer
		{Mo{S}{$_2$}}}.
	\newblock \emph{\bibinfo{journal}{Nano Letters}} \textbf{\bibinfo{volume}{13}},
	\bibinfo{pages}{3626} (\bibinfo{year}{2013}).
	
	\bibitem{Steinhoff15}
	\bibinfo{author}{Steinhoff, A.} \emph{et~al.}
	\newblock \bibinfo{title}{Efficient excitonic photoluminescence in direct and
		indirect band gap monolayer {M}o{S}$_2$}.
	\newblock \emph{\bibinfo{journal}{Nano Letters}} \textbf{\bibinfo{volume}{15}},
	\bibinfo{pages}{6841} (\bibinfo{year}{2015}).
	
	\bibitem{He16}
	\bibinfo{author}{He, X.} \emph{et~al.}
	\newblock \bibinfo{title}{Strain engineering in monolayer {WS}$_2$, {MoS}$_2$
		and the {WS}$_2$/{MoS}$_2$ heterostructure}.
	\newblock \emph{\bibinfo{journal}{Appl. Phys. Lett.}}
	\textbf{\bibinfo{volume}{109}}, \bibinfo{pages}{173105}
	(\bibinfo{year}{2016}).
	
	\bibitem{Chernikov15}
	\bibinfo{author}{Chernikov, A.} \emph{et~al.}
	\newblock \bibinfo{title}{Electrical tuning of exciton binding energies in
		monolayer ${\mathrm{WS}}_{2}$}.
	\newblock \emph{\bibinfo{journal}{Phys. Rev. Lett.}}
	\textbf{\bibinfo{volume}{115}}, \bibinfo{pages}{126802}
	(\bibinfo{year}{2015}).
	
	\bibitem{Ugeda14}
	\bibinfo{author}{Ugeda, M.~M.} \emph{et~al.}
	\newblock \bibinfo{title}{Giant bandgap renormalization and excitonic effects
		in a monolayer transition metal dichalcogenide semiconductor}.
	\newblock \emph{\bibinfo{journal}{Nature Materials}}
	\textbf{\bibinfo{volume}{13}}, \bibinfo{pages}{1091} (\bibinfo{year}{2014}).
	
	\bibitem{Roesner16}
	\bibinfo{author}{R{\"o}sner, M.} \emph{et~al.}
	\newblock \bibinfo{title}{Two-dimensional heterojunctions from nonlocal
		manipulations of the interactions}.
	\newblock \emph{\bibinfo{journal}{Nano Lett.}} \textbf{\bibinfo{volume}{16}},
	\bibinfo{pages}{2322} (\bibinfo{year}{2016}).
	
	\bibitem{Raja17}
	\bibinfo{author}{Raja, A.} \emph{et~al.}
	\newblock \bibinfo{title}{Coulomb engineering of the bandgap and excitons in
		two-dimensional materials}.
	\newblock \emph{\bibinfo{journal}{Nature Commun.}}
	\textbf{\bibinfo{volume}{8}}, \bibinfo{pages}{15251} (\bibinfo{year}{2017}).
	
	\bibitem{Florian17}
	\bibinfo{author}{Florian, M.} \emph{et~al.}
	\newblock \bibinfo{title}{The dielectric impact of layer distances on exciton
		and trion binding energies in van der Waals heterostructures}.
	\newblock \emph{\bibinfo{journal}{arXiv:1712.05607}}  (\bibinfo{year}{2017}).
	
	\bibitem{Plechinger15}
	\bibinfo{author}{Plechinger, G.} \emph{et~al.}
	\newblock \bibinfo{title}{Identification of excitons, trions and biexcitons in
		single-layer {WS}$_2$}.
	\newblock \emph{\bibinfo{journal}{Phys. Status Solidi RRL}}
	\textbf{\bibinfo{volume}{9}}, \bibinfo{pages}{4547--461}
	(\bibinfo{year}{2015}).
	
	\bibitem{IlesSmith16}
	\bibinfo{author}{{{I}les-{S}mith}, J.} \& \bibinfo{author}{Nazir, A.}
	\newblock \bibinfo{title}{Quantum correlations of light and matter through
		environmental transitions}.
	\newblock \emph{\bibinfo{journal}{Optica}} \textbf{\bibinfo{volume}{3}},
	\bibinfo{pages}{207} (\bibinfo{year}{2016}).
	
	\bibitem{Chovan08}
	\bibinfo{author}{Chovan, J.}, \bibinfo{author}{Perakis, I.~E.},
	\bibinfo{author}{Ceccarelli, S.} \& \bibinfo{author}{Lidzey, D.~G.}
	\newblock \bibinfo{title}{Controlling the interactions between polaritons and
		molecular vibrations in strongly coupled organic semiconductor
		microcavities}.
	\newblock \emph{\bibinfo{journal}{Phys. Rev. B}} \textbf{\bibinfo{volume}{78}},
	\bibinfo{pages}{045320} (\bibinfo{year}{2008}).
	
	\bibitem{Michetti09}
	\bibinfo{author}{Michetti, P.} \& \bibinfo{author}{{{L}a {R}occa}, G.~C.}
	\newblock \bibinfo{title}{Exciton-phonon scattering and photoexcitation
		dynamics in J-aggregate microcavities}.
	\newblock \emph{\bibinfo{journal}{Phys. Rev. B}} \textbf{\bibinfo{volume}{79}},
	\bibinfo{pages}{035325} (\bibinfo{year}{2009}).
	
	\bibitem{MolinaSanchez11}
	\bibinfo{author}{{M}olina {S}anchez, A.} \& \bibinfo{author}{Wirtz, L.}
	\newblock \bibinfo{title}{Phonons in sinlge-layer and few-layer {MoS$_2$} and
		{WS$_2$}}.
	\newblock \emph{\bibinfo{journal}{Phys. Rev. B}} \textbf{\bibinfo{volume}{84}},
	\bibinfo{pages}{155413} (\bibinfo{year}{2011}).
	
	\bibitem{Jones15}
	\bibinfo{author}{Jones, A.~M.} \emph{et~al.}
	\newblock \bibinfo{title}{Excitonic luminescence upconversion in a
		two-dimensional semiconductor}.
	\newblock \emph{\bibinfo{journal}{Nature Physics}}
	\textbf{\bibinfo{volume}{12}}, \bibinfo{pages}{323} (\bibinfo{year}{2015}).
	
	\bibitem{deLiberato09}
	\bibinfo{author}{Liberato, S.~D.} \& \bibinfo{author}{Ciuti, C.}
	\newblock \bibinfo{title}{Stimulated scattering and lasing of intersubband
		cavity polaritons}.
	\newblock \emph{\bibinfo{journal}{Phys. Rev. Lett.}}
	\textbf{\bibinfo{volume}{102}}, \bibinfo{pages}{136403}
	(\bibinfo{year}{2009}).
	
	\bibitem{Hunger12}
	\bibinfo{author}{Hunger, D.}, \bibinfo{author}{Deutsch, C.},
	\bibinfo{author}{Barbour, R.~J.}, \bibinfo{author}{Warburton, R.~J.} \&
	\bibinfo{author}{Reichel, J.}
	\newblock \bibinfo{title}{Laser micro-fabrication of concave, low-roughness
		features in silica}.
	\newblock \emph{\bibinfo{journal}{AIP Advances}} \textbf{\bibinfo{volume}{2}},
	\bibinfo{pages}{012119} (\bibinfo{year}{2012}).
	
	\bibitem{Kaupp16}
	\bibinfo{author}{Kaupp, H.} \emph{et~al.}
	\newblock \bibinfo{title}{Purcell-enhanced single-photon emission from
		nitrogen-vacancy centers coupled to a tunable microcavity}.
	\newblock \emph{\bibinfo{journal}{Phys. Rev. Applied}}
	\textbf{\bibinfo{volume}{6}}, \bibinfo{pages}{054010} (\bibinfo{year}{2016}).
	
\end{thebibliography}

\begin{thebibliography}{10}
	\bibitem{Rooney17}
	\bibinfo{author}{Rooney, A.~P.} \emph{et~al.}
	\newblock \bibinfo{title}{Observing imperfection in atomic interfaces for van
		der waals heterostructures}.
	\newblock \emph{\bibinfo{journal}{Nano Lett.}} \textbf{\bibinfo{volume}{17}},
	\bibinfo{pages}{5222} (\bibinfo{year}{2017}).
		\end{thebibliography}
\end{document}